\documentstyle [12pt] {article}
\include{dspace12}
\begin{document}
\rightmargin -2.75cm
\textheight 23.0cm
\topmargin -0.5in
\baselineskip 16pt
\parskip 18pt
\parindent 30pt

\title{ \begin{flushright}\vspace{-.5in}
 \normalsize TIFR-TH-93-26\\
\normalsize PRL-TH-93/9\\
 \normalsize May 1993
\end{flushright}
\large \bf Mass Limits of Invisibly Decaying Higgs Particles
\mbox{from the LEP Data}}
\author{\vspace{.2cm}Biswajoy Brahmachari, Anjan S. Joshipura,  Saurabh D.
Rindani\\  Theory Group, Physical Research Laboratory\\
Navrangpura, Ahmedabad 380 009, India \vspace{.4cm} \\
\vspace{.2cm}D.P. Roy\\Theory Group,
Tata Institute of Fundamental Research\\ Homi Bhabha Road,
Bombay 400 005, India\vspace{.4cm}\\
\vspace{.2cm}Sridhar K.\\ Theory Division, CERN,
CH 1211 Geneva 23, Switzerland}
\date{}
\pagestyle{empty}
\pagenumbering{arabic}
\baselineskip 24pt

\maketitle
\baselineskip 16pt
\parskip 16pt
\pagestyle{plain}
\newcommand{\ra}{\rightarrow}
\newcommand{\be}{\begin{equation}}
\newcommand{\ee}{\end{equation}}
\begin{abstract}
In the Majoron models the $SU(2)$ doublet Higgs can decay invisibly into a
Majoron pair via its mixing with a singlet.  An analysis of the LEP data
shows the invisible decay mode to be more visible than the SM decay.  For
these models, the dominantly doublet Higgs $H$ is shown to have a mass
limit within $\pm 6$ GeV of the SM limit irrespective of the model
parameters.  But the dominantly singlet one $S$ can be arbitrarily light
for sufficiently small mixing angle.
\end{abstract}
\newpage
The $e^+e^-$ collider  LEP at CERN has helped in establishing
many of the crucial aspects of the standard model (SM). It has
however not yet found the Higgs boson which plays a key role in
the breaking of the $SU(2)\times U(1)$ invariance of SM. The
latest limit [1] on its mass $M_H$ coming from the recent LEP data
is nearly 60 GeV.

The published ALEPH data from LEP have also been used to constrain
parameters of
extensions of SM [2]. Many extensions such as multi-Higgs or
supersymmetric models [3] contain additional Higgs doublets
which change the couplings of the SM Higgs to the $Z$ and fermions
through mixing. This results in somewhat relaxed limits on the
Higgs mass [2] compared to SM. There exists some extensions
which are however qualitatively different. These extensions, to
be generically called Majoron models [4], are characterized by
the presence of a Goldstone boson. The couplings of this
Goldstone boson to the Higgs are not required in these models to
be small on any theoretical or phenomenological grounds. As a
consequence, the physical Higgs could decay into an invisible
channel containing a Majoron pair [5-8].  Although the importance of
extending the Higgs search at LEP to the invisible decay channel has been
repeatedly emphasised over the past decade [5-8], there has been no
quantitative effort in this direction so far.  The present work is devoted
to this exercise.  We analyse the published ALEPH data [2] for the
invisible as well as the SM decay modes of a Higgs particle using a parton
level MC event generator.  The missing energy channels with 1 or 2 jets
which are the most viable channels for the SM Higgs search, are seen to be
even more viable for the invisibly decaying Higgs particle.  Thus the data
gives even stronger Higgs mass limits in the latter case compared to the
former, as we shall see below.

The key features shared by all the Majoron models $MM$ [4-11] is a
spontaneously broken global $U(1)$ symmetry and a complex $SU(2)
\times U(1)$ singlet scalar field $\eta$ transforming
non-trivially under the
global $U(1)$ [14]. This $U(1)$ could have  different
physical meanings in different models, e.g.,  total lepton number
[4], combination of lepton numbers [9], or a more general family
symmetry [10]. Moreover, the model may
contain more than one Higgs doublet [11], Higgs triplets [6,13],
or more singlets as in the supersymmetric model with spontaneous
$R$ parity breaking [8]. But irrespective of these details, all
models will contain at least one more real scalar ($\eta
_R\equiv {\rm Re}\;\eta/\sqrt{2}$)
mixing with the neutral component $\phi _R$ of the conventional
Higgs doublet. In the simplest case of only  one doublet and
one singlet, the physical states are given by
\begin{equation}
H=\cos \theta \phi_R + \sin\theta\eta_R
\end{equation}
\begin{equation}
S=-\sin \theta \phi_R + \cos\theta\eta_R,
\end{equation}
where $\theta$ is a mixing angle.  Without any loss of generality $\theta$
can be chosen to lie in the range $0 -45^\circ$.  With this choice $H$ and
$S$ have dominant doublet and singlet components respectively.  We shall
follow this choice.

The spontaneous breaking of the global $U(1)$ generates a
Majoron $J\equiv {\rm Im}\;\eta/\sqrt{2}$ in all the above
models. This gets coupled to massive Higgs through a quartic
term $\delta\; \phi^\dagger \phi\;\eta^\dagger\eta$ in the
scalar potential. Such a coupling cannot be easily forbidden by
any symmetry [15] and leads to the decays $H\ra JJ$ and
 $S\ra JJ$ which
dominate over the conventional $H\ra
b\overline{b}$ and $S\ra b\overline{b}$ decays for a large
range in the relevant parameters [5].
Thus the Higgs bosons in these models are expected
to decay dominantly into an invisible channel and the
experimental search should take this channel into account [16].

The SM Higgs is assumed to be  produced at LEP through the Bjorken process
$e^+e^- \ra Z^*H\ra f\overline{f}H$ [3] and is
searched for [1,2] by means of the signals (A) $l^+l^-\;\;+\;\;$ one
or two jets, and (B) one or two jets $+$ missing energy. The
signal in (A) would be produced by $Z^*\ra l\overline{l},
H\ra b\overline{b}$, while (B) is generated when
$Z^*\ra \nu\overline{\nu}$ and $H\ra
b\overline{b}$. In Majoron models, the signal in (A) is expected
to get diluted [5,11] due to a considerable reduction in the
branching ratio for $H\ra f\overline{f}$. A similar
dilution would also occur in (B), but in this case a new mode
namely $Z^*\ra q\overline{q}$ and $H\ra JJ$ also
contribute to the same signal. Indeed this new mode more than compensates
for the loss of the SM decay signals, thanks to the larger branching
fraction of $Z^\star \ra q\overline{q}$ relative to
$\nu\overline{\nu}$ and $\ell\overline{\ell}$.  In the following, we derive
general limits on the Higgs masses including both the
contributions (A) and (B).

The spin averaged matrix element square for the Bjorken process in the SM
is
\be
\begin{array}{l}
\displaystyle |\overline{M}|^2_{e^+e^- \ra Z^\star H \ra f\overline{f}
H} = {2(4\pi\alpha)^3 M^2_Z C_f \over \sin^6 \theta_W \cos^6 \theta_W}\\[3ex]
\displaystyle \qquad\qquad\qquad\qquad \quad \left[ {C_1 (p_+ \cdot q_-)
(p_- \cdot q_+) + C_2 (p_+ \cdot q_+) (p_-
\cdot q_-) \over \left\{(s-M^2_Z)^2 + M^2_Z \Gamma^2_Z\right\} \left\{(s_1
- M^2_Z) + M^2_Z \Gamma^2_Z\right\}} \right] \nonumber
\end{array}
\ee
with
\be
C_{1,2} = (C^{e^2}_V + C^{e^2}_A) (C^{f^2}_V + C^{f^2}_A) \pm 4C^e_V C^e_A
C^f_V C^f_A, \nonumber
\ee
where $p_\pm$ and $q_\pm$ are the $e^\pm$ and $\overline{f},f$ momenta;
and $s_1$ stands for the virtual $Z^\star$ mass.  The colour factor $C_f$
is $3(1)$ for quarks (leptons); and the vector and axial-vector couplings
are defined in terms of the weak isospin $T_3$ and electric charge $Q$ as
\be
C^f_V = T^f_3 - 2Q^f \sin^2 \theta_W, ~~~C_A^f = T^f_3. \nonumber
\ee
For the massive $b$ quark there is an additional term in the numerator of
the square bracket of (3), i.e.
$$
\begin{array}{l}
m^2_b (C^{e^2}_V + C^{e^2}_A) \Bigg\{(C^{b^2}_V - C^{b^2}_A) {s \over 2} +
C^{b^2}_A {(s_1 - 2M^2_Z) \over M^4_Z} \cdot \\[2mm]
\qquad \qquad \qquad \quad \cdot (2p_+ \cdot (q_+ + q_-) p_- \cdot (q_+ +
q_-) - s s_1/2)\Bigg\}
\end{array}
\eqno (3a)
$$

Due to mixing, eqs.(1-2), the production of $H(S)$ will be reduced by
$\cos^2\theta$ ($\sin^2\theta$) compared to eq.(3). Eq.(3) can
be used to obtain the expected number of events consistent with
the experimental cuts imposed to reduce the background.

We shall work with the published ALEPH data [2], because this is the only
LEP data we could find which explicitly describes all the experimental
cuts.  This will be required for our quantitative MC analysis.  The data
sample corresponds to a little over 185000 hadronic $Z$ events, spread
over a CM energy range of 88.2 -- 94.2 GeV at intervals of 1 GeV.  The
expected number of events for the Bjorken process (3) at each energy is
obtained by multiplying the corresponding number of hadronic $Z$ events by
the ratio of the two cross-sections.  Thus the effect of initial state
radiation factors out from the normalisation [17].  What remains
unaccounted for is a slight reduction of the final state particle
$(Hf\overline{f})$ momenta due to the ISR.  This is negligible for our
purpose, however, since only 2\% of the LEP events have an ISR photon
energy exceeding 2 GeV [18].

The dominant channels for the Higgs signal in the SM $(HZ^\star \ra
b\overline{b} \nu\overline{\nu})$ as well as the Majoron models MM
$(HZ^\star \ra JJ q\overline{q})$ are the missing energy channels
containing 1 or 2 jets.  Therefore we shall concentrate on the
experimental data and cuts of ALEPH [2] in these channels.  The two
channels are separated by defining two hemispheres with respect to the
thrust axis.  Events with total energy deposit $< 2$ GeV in one of the
hemispheres constitute the 1) monojet channel while the remainder
constitute the 2) acoplanar jets channel.

Since the 2nd channel dominates the signal for most of the Higgs mass
range, let us discuss it in some detail.  In our parton level MC
simulation it corresponds to events, where the angle between the two
quarks $\theta_{jj} > 90^\circ$ and the softer quark has an energy $> 2$
GeV.  Table I summarises the effects of the ALEPH cuts on a 50 GeV Higgs
signal for the SM and MM decays.  The corresponding results for the ALEPH
simulation of a 50 GeV SM Higgs signal are also shown along with their
data.  To start with, the missing energy $E\!\!\!\!/$ cut is implimented
through a visible mass cut $M_{jj} < 70$ GeV.  The low angle cut requires
the energy coming out within $12^\circ$ of the beam axis to be $< 3$ GeV
and that beyond $30^\circ$ of the beam axis to be $> 60$\% of the visible
energy.  It removes events with jets close to the beam pipe, where
measurement errors can simulate a $E\!\!\!\!/$.  The acollinearity cut
removes the $Z \ra q\overline{q}$ and $\tau^+\tau^-$ background where the
$E\!\!\!\!/$ can be due to fluctuation (including escaping $\nu$) of one
or both the jets.  The $\tan\alpha > 4$ cut for the missing momentum
$\vec{p\!\!\!/}$, making an angle $\alpha$ with the beam axis, removes the
$E\!\!\!\!/$ background from ISR and $e^+e^- \ra (e^+e^-)\gamma\gamma$
processes.  The isolation cut removes $E\!\!\!\!/$ background from the
fluctuation of any one of the jets.  The acoplanarity cut for a 3-jet like
events removes that due to fluctuation of one or more of these jets.  This
is analogous to the acollinearity cut for the 2-jet events.  The
acoplanarity cut for the 2-jet events removes the $E\!\!\!\!/$ background
arising from ISR along with the fluctuation of one or both the jets.  The
remaining few events are the residual $\gamma\gamma$ events, which are
removed by the total $p_T$ cut.  The last line shows that the original
visible mass cut is dispensible, since all the events are removed even
without it.

The cuts are seen to have no strong effect on the Higgs signal for either
the SM or the MM, since they naturally simulate a large missing energy
which is neither tied up to the beam nor the jet directions.  Comparing
our parton level MC simulation for the SM Higgs signal with the full MC
simulation of ALEPH one sees an agreement to within 10\% for any
combination of the cuts.  Since a 10\% variation in the signal corresponds
to a $< 1$ GeV charge in $M_H$, we expect the $M_H$ limit from the parton
level MC to be reliable to within 1 GeV.  The overall efficiency factor in
the two cases of course agree at the level of $\sim 2$\%.  We have checked
that the agreement continues to be good to $\sim 10$\% for $M_H \geq 20$
GeV [19].  For $M_H < 20$ GeV, the $H \ra \tau^+\tau^-,D\overline{D}$ decay
modes become important; so that there is an appreciable loss of efficiency
due to the ALEPH requirement of at least 5 good tracks.  Although one
could incorporate this into the parton level MC, we felt it unnecessary to
extend our SM Higgs analysis to this region.  Comparing the Higgs signals
for the SM and MM decays one again sees that the effect of the cuts are
very similar in the two cases.  This is to some extent accidential; $M_H =
50$ GeV corresponds to a peak value of $M_{Z^\star} \simeq 40$ GeV, so
that the decay quark jets have very similar kinematics for the two cases.
At lower $M_H$, the efficiency factor for the MM decay is somewhat lower
than the SM decay (Fig. 1).  This is partly due to the 70 GeV mass cut
which affects the MM decay signal as $M_{Z^\star}$ increases with
decreasing $M_H$.  For the same reason, however, the decay quark jets are
expected to be hard and hence satisfy the requirement of $\geq 5$ good
tracks automatically.  Consequently our parton level MC result should hold
even at low Higgs mass for the MM decay.

Fig. 1 shows the expected number of signal events as a function of the
Higgs mass for the SM and MM decays.  The normalisation corresponds to the
production cross-section from (3) -- i.e. it corresponds to the limit
$\theta \ra 0$ when $H$ becomes essentially a doublet.  The event rates
are shown both before and after the experimental cuts.  The contributions
of the two channels to the latter rate are also shown separately.  As
expected the SM signal is dominated by the monojet and acoplanar jets
contributions at small and large $M_H$ respectively.  The MM signal is
dominated by the acoplanar jets throughout the $M_H$ range of interest;
but the importance of the monojet contribution increases with increasing
$M_H$ (i.e. decreasing $M_{Z^\star}$).  Comparing the SM and MM decay
signals we see that the latter is larger by a factor of $2 \ra 3$ for $m_H
= 20 \ra 50$ GeV.  This is due to the larger branching fraction of
$Z^\star \ra q\overline{q}$ relative to $\nu\overline{\nu}$, which remains
largely unaffected by the cuts.  The SM decay signal can be increased a
little by including contributions from other channels, notably $Z^\star
\ra \ell^+\ell^-$.  The crosses denote the resulting signal taken from
[2].  Still the size of the SM decay signal remains small relative to the
MM decay.  This clearly demonstrates that an invisible decay mode of a
Higgs particle would be more visible at LEP compared to the SM decay.  The
95\% CL limits on $M_H$, corresponding to 3 signal events, are 48 and 54
GeV for the SM and MM decays respectively.  It may be noted here that the
latest $M_H$ limit of $\sim 60$ GeV for the SM decay [1] would roughly
correspond to a 65 GeV limit for the MM decay.

In general one expects both the SM and MM decays to occur with a relative
branching ratio $r$ say.  Moreover, the physical Higgs particles $H$
and $S$ are expected to be combinations of the doublet and singlet fields.
Thus the expected size of the Higgs signal is in general
\be
\begin{array}{l}
N_{\rm exp} = \cos^2\theta \left\{N_{SM} (M_H) {r_H \over 1 +
r_H} + N_{MM} (M_H) {1 \over 1 + r_H}\right\} \\[2mm]
\qquad \quad + \sin^2\theta \left\{ N_{SM} (M_S) {r_S \over 1 +
r_S} + N_{MM} (M_S) {1 \over 1 + r_S}\right\}, \nonumber
\end{array}
\ee
where $N_{SM}$ and $N_{MM}$ correspond to the crosses of Fig. 1a and the
solid line of Fig. 1b respectively.  One can get independent limits on
$M_H$ and $M_S$ by assuming that only one of them contributes to the
signal.  These will be somewhat weaker than the joint limit of course.
Fig. 2 shows the independent limits on $M_H$ and $M_S$ for the extreme
values of $r_{H,S} = 0$ and $\infty$ as a functions of the mixing
angle $\theta$.  Thus the 2 bands represent the lower limits of $M_H$ and
$M_S$ over the entire parameter space.  Note that the limit of $M_H$,
representing the physical Higgs particle with larger doublet component, is
remarkably stable vis a vis the SM limit.  For small $\theta$, where $H$
is dominated by the doublet component, the limit increases from 48 to 54
GeV as the branching fraction for the $MM$ decay increases from 0 to 1 as
expected from Fig. 1.  Increasing the mixing angle $\theta$ to its maximal
value of $45^\circ$ decreases the production rate by a factor of 2 and
correspondingly the $M_H$ limit by $\sim 6$ GeV.  Thus
\be
M^{\rm lim.}_H = M^{\rm lim.}_{SM} \pm 6~{\rm GeV} \nonumber
\ee
for the entire parameter space.  Again this correlation should hold for
the recent LEP data [1] as well.  The $M_S$ limits coincide with $M_H$ at
$\theta = 45^\circ$ as expected from (6); but goes down steadily with
$\theta$ (i.e. the $SZZ$ coupling).  Thus the $M_S$ limit goes down to
$\sim 10$ GeV for $\theta \sim 10^\circ$, below which there is no $M_S$
limit from the published ALEPH data [2].  It is easy to translate this
into a $M_S$ mass limit of $\sim 10$ GeV for $\theta \sim 5^\circ$ for the
recent data [1].  Recently some of the above points have been discussed at
a qualitative level in [20].

Finally, Fig. 3 shows the joint limit on $M_H$ and $M_S$ from (6) for
$r_{H,S} = 0$, where the invisible decay mode dominates for both the
Higgs particles.  The limits are shown for 3 representative values of the
mixing angle.  The corresponding limits for $r_{H,S} = \infty$ are
essentially given by parallel curves shifted to the left by $\sim 6$ GeV.

We shall conclude by relating the relative branching ratios $r_{H,S}$
to the underlying model parameters.  In the simplest case discussed above,
the Higgs sector of the model with $\phi$ and $\eta$
fields contains [5] one more independent parameter, $\tan\beta
\equiv
\langle\phi\rangle/\langle\eta\rangle$, in
addition to $M_H$, $M_S$ and $\theta$.  The relative branching ratios are
given by
\be
r_H\approx\frac{1}{12}\left(\frac{m_b}{M_H}\right)
^2\cot^2\theta\cot^2\beta \left(1-4\frac{m_b^2}{M_H^2}\right)
 ^{ 3/2 }
\ee
\be
r_S\approx\frac{1}{12}\left(\frac{m_b}{M_S}\right)
^2\tan^2\theta\cot^2\beta \left(1-4\frac{m_b^2}{M_S^2}\right)
 ^{3/2}.
\ee
The actual value of $\theta$ as well as $\tan\beta$ is detemined
by the scale of the global $U(1)$ breaking relative to the
$SU(2)\times U(1)$ breaking scale, as well as by the quartic
couplings in the Higgs potential. Typical expectations would
thus be $\tan\beta\approx O(1)$ and $\cos\theta\approx O(1)$, if
the two scales coincide. In the event of $\langle\eta\rangle
>\!> \langle\phi\rangle$, $\tan\beta$ is very small and the
Higgs decay to a Majoron pair is supressed. In the converse
limit of $\langle\eta\rangle <\!< \langle\phi\rangle$, the Higgs
mainly decay [11] to  Majoron pairs if the coefficient of the
quartic term $\phi^\dagger\phi\;\eta^\dagger\eta$ in the
potential is $O(1)$.

In the foregoing discussions, we have assumed that the Higgs
sector of the Majoron model contains a doublet and a singlet.
Some models [8,11] do require additional scalar fields either as
doublets or singlets. The Higgs fields in these models will not
be given by  simple expressions like eqs.(1) and (2) containing
only one mixing angle. The limits on the various Higgs masses
would be different in these models, since the production cross
sections would now be suppressed by different amounts and
$N_{exp}$ would change accordingly. But generically, one should
be able to derive strong constraints in these models as well.
Specifically, in the absence of large mixing angles, the limit of
$M_H>54$ would be applicable to these models as well, if Higgs
decays mainly into the invisible channel.

In summary, we have considered here limits on the Higgs mass in
Majoron models which contain the interesting possibility of the
decay of Higgs bosons to an invisible channel. A minimal models of
this type contains two Higgs scalars $H$ and $S$ which are
predominantly $SU(2)\times U(1)$ doublet and singlet,
respectively. We have shown that the LEP data imply a stringent
limit on the mass of the $H$. Specifically,  the limit on $M_H$
could be better than in SM; and in any case it should lie
within $\pm 6$ GeV of the SM limit. In contrast, the predominantly singlet
Higgs $S$ could be even lighter than 10 GeV. We have used in this work
only the published ALEPH data [2].  A similar analysis of the more recent
data [1] would evidently strengthen these results [21].

B.B would like to acknowledge the hospitality and financial support of
Tata Institute of Fundamental Research during the course of this work.
D.P.R acknowledges discussion with T. Aziz, A. Gurtu and K. Sudhakar on
the LEP data.

\newpage

\centerline{ \large \bf  References}

\noindent [1] J.F. Grivaz, Orsay report no. LAL 92-59 (1992).\\
\noindent [2] D. Decamp et al., ALEPH Collaboration, Phys. Rep.
{\bf 216}, 253 (1992); CERN report CERN-PPE-91-19 (1991).\\
\noindent [3] J. Gunion et al., {\it The Higgs Hunter's Guide},
(Addison Wesley, Menlo Park, 1991)\\
\noindent [4] Y. Chikashige, R.N. Mohapatra and R.D. Peccei, Phys. Lett.
{\bf 98B}, 265 (1980).\\
\noindent [5] A.S. Joshipura and S.D. Rindani, Phys. Rev. Lett. {\bf 69},
3269 (1992). \\
\noindent [6] R.E. Schrock and M. Suzuki, Phys. Lett. {\bf 10B}, 250
(1982); L.F. Li, Y. Liu and L. Wolfenstein, Phys. Lett. {\bf
159B}, 45 (1985).\\
\noindent [7] E.D. Carlson and L.B. Hall, Phys. Rev.
{\bf D 40} 3187 (1985); G. Jungman and M.A. Luty, Nucl. Phys. {\bf
B361}, 24 (1991).\\
\noindent [8] J.C. Romao, F. de Campos and J.W.F. Valle, Phys.
Lett. {\bf B292}, 329 (1992).\\
\noindent [9] A.S. Joshipura and S.D. Rindani, Phys. Rev. {\bf D 46}, 3000
(1992).\\
\noindent [10] F. Wilczek, Phys. Rev. Lett. {\bf 49}, 1549
(1982); D.B. Reiss, Phys. Lett. {\bf 115B}. 217 (1982).\\
\noindent [11] A.S. Joshipura and J.W.F. Valle, CERN report
CERN-TH. 6652/92, Nucl. Phys. B (to be published).\\
\noindent [12] G. Gelmini and M. Roncadelli, Phys. Lett. {\bf
B99}, 411 (1981).\\
\noindent [13] A.S. Joshipura, Int. J. Mod. Phys. {\bf A7}, 2021
(1992); K. Choi and S. Santamaria, Phys. Lett. {\bf B267}, 504 (1991).\\
\noindent [14] The triplet Majoron model [12] is an exception to this.
But this by itself is ruled out by the LEP data on invisible $Z$
width unless an additional singlet [13] is added.\\
\noindent [15] This is not true in the supersymmetric extension of the
singlet Majoron model where this coupling is forbidden by the
combined requirement of supersymmetry and lepton number conservation.\\
\noindent [16] The same considerations hold even if the global symmetry is
larger than $U(1)$, as e.g. the dark Higgs model of Bjorken.  J.D.
Bjorken, Invited talk at the Symp. on the SSC Laboratory, Corpus Chrisli,
Texas, 13-17 Oct. 1991, SLAC-PUB-5673 (1991).\\
\noindent [17] F.A. Berends and R. Kleiss, Nucl. Phys. {\bf B260}, 32
(1985).\\
\noindent [18] A more detailed $MC$ analysis including ISR along with jet
fragmentation and covering the recent L3 data is in progress.  A. Gurtu,
K. Sudhakar and D.P. Roy (work in progress).\\
\noindent [19] One can easily check this by comparing the solid line of
Fig. 1a) with the corresponding curve of ref. [2].\\
\noindent [20] A. Lopez-Fernandez et al., Valencia Preprint FTUV/93-19,
April 1993.\\
\noindent [21] On completion of this work came to learn that the ALEPH
group is finalising the analysis of the full data sample in terms of an
invisibly decaying Higgs signal.  A part of their preliminary result has
been reported in ``Searches for Signals of Supersymmetry'', by J.F.
Grivaz, Invited talk at the XXVIIIth Recontre de Moriond, 13-20 March
1993, Orsay preprint LAL 93-11, April 1993.  We are grateful to Prof.
Grivaz for this information.\\
\newpage
\begin{description}

\item[\rm Table I.] Effect of ALEPH cuts in the acoplanar jets channel on
the parton level MC simulation of a 50 GeV Higgs signal for the Standard
Model and Majoron Model (Invisible Decay).  The corresponding results for
the ALEPH simulation of a 50 GeV SM Higgs signal are also shown along with
their Data.

\end{description}
$$
\begin{small}
\begin{tabular}{|l|c|c c c|}
\hline
\multicolumn{1}{|l|}{\qquad \qquad Cut} &
\multicolumn{1}{c|}{Data} &
\multicolumn{1}{c}{Efficiency (\%)} &
\multicolumn{1}{c}{of a 50 GeV} &
\multicolumn{1}{c|}{Higgs signal}\\
& No. of Events & SM (ALEPH) & SM & MM\\
\hline
Mass Cut & 11,865 & 99.4 & 100 & 100\\
$M_{\rm vis.} < 70$ GeV &&&&\\
\hline
Low Angle Cut & 5,018 & 90.4 & ~81 & ~77\\
$E_{12} < 3$ GeV, $E^{30} < .6 E_{\rm vis.}$ &&&&\\
\hline
Acollinearity & ~~~305 & 83.2 & ~72 & ~71\\
$\theta_{jj} < 165^\circ$ &&&&\\
\hline
Low Angle Cut for $\vec{p\!\!\!/}$ & ~~~155 & 78.8 & ~70 & 69.5\\
${\rm Tan}~\alpha > .4$ &&&&\\
\hline
Isolation of $\vec{p\!\!\!/}$ & ~~~~73 & 75.1 & ~70 & 68.6\\
$E_{\rm cone} < 3$ GeV &&&&\\
\hline
Acoplanarity for &&&&\\
3 jet events & ~~~~19 & 71.2 & ~70 & 68.6\\
$S = \displaystyle \sum^3_{i=1} \theta_i < 350^\circ$ &&&&\\
if $\theta^{\rm min}_i > 40^\circ$ &&&&\\
\hline
Acoplanarity & ~~~~~7 & 67.7 & 66.5 & 66.4\\
$\phi_{jj} < 175^\circ$ &&&&\\
\hline
$\gamma\gamma$ Bg. Cut & ~~~~~0 & 67.7 & 66.5 & 66.4\\
$\displaystyle |\sum \vec{p_T}| > .05 E_{CM}$ &&&&\\
if $M_{\rm vis} < 25$ GeV &&&&\\
\hline
Without Mass Cut & ~~~~~0 & 67.8 & 66.5 & 66.4\\
\hline
\end{tabular}
\end{small}
$$
\newpage
\begin{center}
\underbar{\bf Figure Captions}\\
\end{center}

\begin{description}

\item[\rm Fig. 1.] The expected Higgs signals for the SM and MM (invisible)
decay modes corresponding to the published ALEPH data [2].

\item[\rm Fig. 2.] Mass limits for $H$ and $S$ shown as functions of the
mixing angle for the two limits corresponding to predominant MM $(r = 0)$
and SM $(r = \infty)$ decays.

\item[\rm Fig. 3.] Joint mass limit for $H$ and $S$ shown for $r_{H,S} = 0$
and 3 representative values of the mixing angle $\theta$.  The region to
the right of the curves are allowed.

\end{description}
\end{document}